\documentclass[]{interact}

\usepackage{epstopdf}
\usepackage[caption=false]{subfig}
\usepackage[boxruled,linesnumbered]{algorithm2e}

\usepackage[numbers,sort&compress]{natbib}
\bibpunct[, ]{[}{]}{,}{n}{,}{,}
\renewcommand\bibfont{\fontsize{10}{12}\selectfont}
\usepackage{amssymb}
\usepackage{amsmath}
\usepackage{amsthm}
\usepackage[T1]{fontenc}
\usepackage{eucal}

\usepackage{newtxtext,newtxmath}

\theoremstyle{plain}

\theoremstyle{definition}

\theoremstyle{remark}

\def\boxit#1{%
  \smash{\fboxsep=0pt\llap{\rlap{\fbox{\strut\makebox[#1]{}}}~}}\ignorespaces
}
\def\boxit#1{%
  \smash{\fboxsep=0pt\llap{\rlap{\fbox{\strut\makebox[#1]{}}}~}}\ignorespaces
}

\let\oldnl\nl
\newcommand{\nonl}{\renewcommand{\nl}{\let\nl\oldnl}}
\begin{document}

\title{Filtering ASVs/OTUs via Mutual Information-Based Microbiome Network Analysis}

\author{
\name{Elham Bayat Mokhtari\textsuperscript{a,b}\thanks{\textsuperscript{a}Vaccine Research Center, National Institute of Allergy and Infectious Diseases, National Institutes of Health, Bethesda, MD 20892, USA. CONTACT: Elham Bayat Mokhtari. Email: ellie.bayatmokhtari@nih.gov} and Benjamin Ridenhour\textsuperscript{b}}
\affil{\textsuperscript{b}Department of Mathematics and Statistical Science, University of Idaho, Moscow, ID, USA}
}

\maketitle
Word count: 4734\\

\begin{abstract}
Microbial communities are widely studied using high-throughput sequencing techniques, such as 16S rRNA gene sequencing. These techniques have attracted biologists as they offer powerful tools to explore microbial communities and investigate their patterns of diversity in biological and biomedical samples at remarkable resolution. However, the accuracy of these methods can negatively affected by the presence of contamination. Several studies have recognized that contamination is a common problem in microbial studies and have offered promising computational and laboratory-based approaches to assess and remove contaminants. Here we propose a novel strategy, MI-based (mutual information based) filtering method, which uses information theoretic functionals and graph theory to identify and remove contaminants. We applied MI-based filtering method to a mock community data set and evaluated the amount of information loss due to filtering taxa. We also compared our method to commonly practice traditional filtering methods. In a mock community data set, MI-based filtering approach maintained the true bacteria in the community without significant loss of information. Our results indicate that MI-based filtering method effectively identifies and removes contaminants in microbial communities and hence it can be beneficial as a filtering method to microbiome studies. We believe our filtering method has two advantages over traditional filtering methods. First, it does not required an arbitrary choice of threshold and second, it is able to detect true taxa with low abundance.
\end{abstract}


\section{Introduction}

High-throughput sequencing approaches are some of the most powerful tools for studying and characterizing microbial communities. Bacterial phylogeny and taxonomy can be characterized using marker genes, such as 16S rRNA gene sequences which are present in all bacteria, and it is sufficiently large for informatics and analysis purposes \cite{Janda2007,Patel2001}. However, the potential for contamination which is defined as non-intended introduction of bacteria during sample collection, DNA extraction, and PCR amplification into the sample of interest is high; thus a low signal-to-noise ratio poses a major problem in analyses of such data \cite{Weiss2014,Brooks2015, Salter2014}. Contamination is particularly problematic when studying low yield samples because of significant impacts on results \cite{Salter2014,Weiss2014}. Therefore, it is necessary to identify, minimize, and filter contaminants as a potential source of bias that leads to skew data analysis.

Attempts to experimentally control or eliminate sources of contamination can be challenging if not impossible. To minimize or identify contamination, strategies such as inclusion of negative controls or blanks for every batch of samples and use them through entire extraction, amplification, or library preparations have been suggested \cite{Barton2006,Salter2014}. One of the advantages of sequencing the blanks is the ability to detect and quantify the levels of contamination as well as their the sources. \cite{Minich2018,DeGoffau2018,Pollock2018,Barton2006,Salter2014}. However, including an appropriate negative control is not always easy and in the majority of  microbiome published studies controls have not been included \cite{Hornung2019}. \cite{Weiss2014} and \cite{Salter2014} recommended keeping records of kits and other reagents,  performing technical replicates, and using sample randomization across kits and PCR runs to control measurement error. Some researchers have proposed using mock communities as a positive control during extraction, amplification, and sequencing alongside experimental samples \cite{Brooks2015}. Positive controls are commercially available in the form of defined communities, however their validity for a particular microbiome research is not guaranteed and standardized protocols for designing positive controls might not be available \cite{Hornung2019}. 

None of the above experimental methods are capable of eliminating existing contaminants completely, easily, and reliably in all cases. Therefore, strategies that use the power of bioinformatics and statistical methods to clean sequencing data must be introduced. For example, \cite{Jervis-Bardy2015} identified and removed Operational Taxonomic Units (OTUs) as potential contaminants if they have strong negative correlation with amplicon counts after 16S library preparation. However, in many cases, contaminant OTUs might occur on the host as well as being present as contamination and therefore, this leads to a higher than desired false positive rate. 
Ad hoc methods such as removing genes or taxa with total read count or percentage smaller than or below an empirical threshold across all samples \cite{Robinson2010,Anders2010,Law2014,Sultan2008,xia2018statistical} are easy to implement and relatively common among microbiome studies. However, choosing an appropriate filtering threshold is a complex problem by itself and an arbitrary choice can bias the results. In addition, the impact of taxa or genes is not directly proportional to their numeric abundance and there might be biological signal among rare taxa---or genes---that is of interest; thus removing low abundance taxa could lead to loss of important information.

The \texttt{decontam} package in R introduced by \cite{Davis2018} has been developed to identify contaminants using statistical models. \cite{Davis2018} demonstrated the accuracy of their method to remove contaminants from a data set generated by \cite{Salter2014}. However, a major limitation of \texttt{decontam} is that it assumes contaminants and true signals are distinct from one another, and this assumption is violated in the case of cross-contamination due to sequences from pooled samples. \cite{McKnight2019} developed the R package \texttt{microDecon} which is based on proportions of contaminant OTUs or Amplicon Sequence Variants (ASVs) in blank samples to identify and remove contaminant reads from meta-barcoding data.  They demonstrated that their method is robust to both high and low contamination levels. They also showed that their approach can recover the real community from the contaminant community even with a large overlap between the two. However, similar to \cite{Davis2018}, in case of the existence of cross-contamination, this method is not effective as it assumes a common source of contamination. Recently, \cite{Smirnova2018} introduced the R package \texttt{PERFect} for microbiome filtering using covariance matrices and compared them to traditional filtering procedures. They showed that for a very strong signal, \texttt{PERFect} provides a more effective contaminant reduction when the signal-to-noise ratio is high. A limitation of their methods is that it is skewed toward retaining dominant taxa, however, this is a common limitation among any filtering methods that does not take into account other types of information such as knowledge about blanks or negative controls.

Here, we propose and validate a method to identify and remove non-bacterial signals that are observed due to contamination or sequencing errors in microbiome data. We use the fact that bacteria live in communities where they rely on one another, and their interactions or coexistence are major drivers of microbial community and function. We utilize a graph model to represent and characterize these interactions and/or coexistence by assuming each taxon is a node and pairwise-bacterial associations are edges in this biological network. We use an information theoretic functional to estimate the strength of these interactions and remove isolated taxa that are not informative to the network as potential noise. We apply permutation and bootstrap based hypothesis testing to measure the probability of increase in information loss due to taxa removal is random. We validate our method using the \cite{Brooks2015} mock community data set. Next, we compare the performance of commonly used ad hoc filtering methods with our proposed method. Finally, we apply our method to a gut microbiome data set.

The rest of this paper is organized as follows. In Section 2, we introduce our filtering method using graph models and information loss measurement. Statistical inference based on bootstrap and permutation hypothesis testing is presented in Section 2. Method validation and comparison with traditional filtering methods using \cite{Brooks2015} data set are provided in Section 3 and 4, respectively. Finally we conclude the paper on Section 5.

\section{Materials and Methods}\label{class}

We propose a method to identify and remove contaminated sequence reads from data sets, while accounting for the amount of information loss due to this removal. Note that the proposed method can be applied to both OTU and ASV count tables.   

\subsection{\textbf{Mathematical Definition}}
Here, we define notations which will be frequently used in the following sections. Consider a high-dimensional count matrix where each input represents the count of sequence reads of a taxon, which, for simplicity, we will assume to be a bacterial species or strain. Let $X_{n\times m}$ be a microbial abundance matrix. For each  $i=1,\cdots,n$ and $j=1,\cdots,m$, let $x_{ij}$ be the observed count of the $j$-th taxon in the $i$-th sample and $X_j$ denotes the abundance of the $j$-th taxon across all $n$ samples. Generally, the number of samples is considerably less than the number of taxa, that is $n<<m$.

\subsection{\textbf{The Proposed Method: Network-based Contaminant Identification}}
Graph theory is an important concept in statistics and can be used to describe the relationships between random variables \citep{sun2018,Li2016}. A network (or a graph) is defined as a set of nodes connected by edges \citep{Naqvi2010}. Microbial interactions can be represented as a connectivity network, where nodes correspond to taxa and the edges represent the associations between taxa \citep{Tavakoli2019}. One potential association measure is mutual information (MI) which is a non-directional connectivity measure.  MI was introduced by Shannon in 1948 \citep{Shannon1948} as a measure of statistical dependence between two random variables. Unlike Pearson or Spearman correlation coefficients, the most widely used association measures, that quantify linear and monotonic relationships, respectively, MI can be used to estimate non-linear relationships \citep{Song2012,DIONISIO2004326}.

MI measures the expected reduction in uncertainty about $X$ that results from learning $Y$, or vice versa. This quantity can be formulated as
\begin{align}
    I(X;Y) = H(X)-H(X|Y),
\end{align}
where $H(X)$, known as ``entropy,'' is the average amount of information, or surprise, a variable $X$ has. It  is defined to be
\begin{align}
    H(X) = -\sum_{x\in\mathcal{X}} p(x)\log p(x),
\end{align}
where $p$ is the probability of observing the $i$-th value of the bin measurement data $x_i \in \mathcal{X}$ using partition-based methods such as histograms. The conditional entropy is the uncertainty of $X$ given $Y$ and it is formulated as
\begin{align}
    H(X|Y=y) = -\sum_{x\in\mathcal{X}}\frac{p(x,y)}{p(y)}\log\frac{p(x,y)}{p(y)},\label{condent}
\end{align}
\noindent
where $p(x,y)$ is the joint probability density of measurements $X$ and $Y$. \\
From equation \eqref{condent} we can derive the following identity
\begin{align}
    H(X|Y) = H(X,Y)-H(X),
\end{align}
\noindent
where $H(X,Y) = -\sum_{x\in{\mathcal{X},y\in\mathcal{Y}}}p(x,y)\log p(x,y)$ is the joint entropy which measures the amount of uncertainty in the two random variables $X$ and $Y$ taken together. 

MI possesses the following desirable properties. 
\begin{enumerate}
    \item It is symmetric: $I(X;Y)=I(Y;X)$,
    \item $I(X;Y)\geq 0$, equality holds if and only if the two variables are independent,
    \item $I(X;Y)\leq H(X,Y)$.
\end{enumerate}
In situations where $X$ is uniquely determined by $Y$, knowledge of $Y$ dictates a single possible value of $X$. It then follows that the conditional entropy satisfies $H(X|Y) = 0$ and therefore MI has the maximum value of $I(X;Y)= H(X)$. Moreover, the stronger the relationship between two variables, the greater is the MI. \citet{Kinney2014} proved that MI places the same importance on linear and nonlinear dependence.

Here, we use MI as an association measure and transform it into network adjacencies. A network adjacency $A=(A_{ij})$ satisfies the following conditions:
\begin{enumerate}
    \item $0\leq A_{ij}\leq 1$,
    \item $A_{ij}=A_{ji}$,
    \item $A_{ii}=1$.
\end{enumerate}
For $m$ taxa $X_1,\cdots,X_m$ an adjacency matrix $\mathcal{I}$ is a $m$ by $m$ matrix  where each entry is the amount of information shared between each pair of taxa. We construct our adjacency matrix based on MI by satisfying three above conditions:  1) transformation to $[0,1]$; 2) symmetrization; and, 3) setting diagonal values to 1. It can be easily seen that MI is bounded below by 0 and it is symmetric. However, it is not bounded above by 1 and the diagonals are not equal to $1$ but rather are the entropy of the variable, $H(X)$. To satisfy the above conditions, we divide each entry of the mutual information matrix $\mathcal{I}$ by one of its upper bound which is a joint entropy between each pair of taxa, resulting in adjusted adjacency matrix $\widetilde{\mathcal{I}}$.

Therefore, for each pair of taxa $X_j$ and $X_{j'}$, the adjusted mutual information is calculated as  
\begin{align}
    \widetilde{\mathcal{I}}_{jj'}(X_j;X_{j'}) = \frac{\mathcal{I}_{jj'}{(X_j;X_{j'})}}{H(X_j,X_{j'})},\quad j,j'= 1,\cdots,m\label{adjmi}
\end{align}

The result of this transformation is $m$ by $m$ matrix  $\widetilde{\mathcal{I}}$ where each entry varies between $0$ and $1$. Also, if $j=j'$, then $\mathcal{I}_{jj'}(X_j;X_{j'})=H(X_j)$ and $H(X_j,X_{j'})=H(X_j)$ so $\widetilde{\mathcal{I}}_{jj'}(X_j;X_{j'})=1$. Thus our transformation (\ref{adjmi}) satisfies the conditions of a network adjacency.

In the following subsection, we describe an approach that results in an unweighted adjacency matrix based on adjusted mutual information measure we defined above.

\subsubsection{\textbf{Filtering Using Unweighted Network Adjacency}}

A filtered unweighted network adjacency between taxa $X_j$ and $X_j'$ can be defined by hard thresholding the adjusted mutual information-based adjacency matrix $\widetilde{\mathcal{I}}$ using signum function.
\begin{align}
    \mathcal{I}_{jj'}^*(X_j;X_j') = \begin{cases}
    1 & \text{if\;\; $\widetilde{\mathcal{I}}_{jj'}(X_j;X_j')\geq\tau$}\\
    0 & \text{otherwise}
  \end{cases}, \label{threshold}
\end{align}
where $\tau$ is the hard threshold parameter. Hard thresholding leads to the intuitive concept of taxa connectivity (i.e., a binary variable indicating whether two species do or do not interact), and it is commonly used to construct a sparse covariance matrices \citep{wgcna2005, Sulaimanov2016}.

\subsubsection*{\textbf{Choosing The Threshold $\tau$}}
In many biological networks, hard thresholding of the association adjacency matrix is based on the scale-free criteria (defined below) of a graph and often applied when $m<<n$ \citep{BB509,Albert4947,ZhangHorvath2005}. In other words, it is assumed that the probability that a node is connected with $k$ other nodes (the degree distribution of a network) is characterized by a power-law distribution
\begin{align}
    P(k)\sim k^{-\gamma},
\end{align}
where $k$ is the node degree, and $\gamma$ is some exponent reported in some biological graphs to be $2<\gamma<3$ \citep{Barabasi2004}. We choose the threshold $\tau$ by fitting a linear function $f(k)=-\hat{\gamma}k+\hat{b}$ to the empirical degree distribution in log space and estimating the coefficient of variation, ($R^2$), of the fit. We choose the threshold that results in the highest $R^2$ value . In addition to high $R^2$ values, it is recommended \citep{wgcna2005, Sulaimanov2016} to have a high mean connectivity so that the network contains enough information. We compute the mean degree $\bar{k}$ for each threshold $\tau$, by taking the average over the degree of all nodes. It is expressed as follows
$$\bar{k}=\frac{\sum_{j'=1}^{m}\sum_{j=1}^{m}\mathcal{I}_{jj'}^*}{m}$$
We use mean connectivity as a tie breaker for thresholds that could produce the same $R^2$ value.
Choosing an appropriate threshold which provides us with the highest $R^2$ and a high $\bar{k}$, we build our network based on $\mathcal{I}_{jj'}^*$ and remove isolated nodes (taxa), i.e., nodes that have connectivity degree of 0. Because isolated nodes do not share information with other taxa, we assume they are potential contaminants, and we may remove them without significant loss of information. Conversely, nodes (taxa) that create non-trivial subgraphs (i.e., subgraphs having more than one node) are assumed to be true taxa. 

\subsection{\textbf{Subnetworks With Minimal Information Loss}}

Hidaka et al. \cite{Hidaka2018} proposed a method of searching graph partitions (separations of the vertex set) which leads to the minimal information loss. In another work,  Smirnova et al. \cite{Smirnova2018} proposed a filtering loss measure to remove taxa with insignificant contribution to the total covariance. Inspired by the these ideas, we propose a method to filter taxa in a network based on total mutual information. 

To do this, first we define the connectivity degree $d_j$ of the $j$-th node for $j=1,\cdots,m$ in the weighted graph; this is the sum of the weights on all edges adjacent to node $j$. The formula for connectivity degree $d_j$ is 
\begin{align}
    d_j = \sum_{j'=1}^m \widetilde{\mathcal{I}}_{jj'},
\end{align}
where we take $\widetilde{\mathcal{I}}_{jj'}$ to be the weight on the edge connecting nodes $j$ and $j'$. Next, we sort the connectivity degree $d_j$ in an increasing order. Following this, we remove nodes (taxa) based on sample quantile values of sorted connectivity degrees for all taxa $j=1,\cdots, m$. Finally, we compute the information loss according to the following formula:
\begin{align}
    \Lambda_k = 1- \frac{\|\widetilde{\mathcal{I}}_k'\|_F^2}{\|\widetilde{\mathcal{I}}\|_F^2},
\end{align}
where $\|\cdot\|^2_F$ is the Frobenius norm, sometimes also called the Euclidean norm, $\widetilde{\mathcal{I}}_k'$ is the adjusted mutual information matrix after removing all taxa below the $k^{th}$ quantile. Here, $\|\widetilde{\mathcal{I}}_k'\|^2_F$ represents the total information shared between taxa after removing certain number of taxa.\\  

\subsection{Statistical Inference: Hypothesis Testing}
\subsubsection{\textbf{Hypothesis Testing Using a Permutation Test}}

 In this subsection we present an algorithm based on permutation testing, described in Algorithm 1, inspired by \citet{Francois2006} to compare the difference in information loss due to various quantile removal. Let $q_1,\cdots,q_{\ell}$ be the quantile values. We are interested in testing if the information loss by removing the taxa with degree less than $q_{k}$ is significantly different from removing taxa with degree less than $q_{k+1}$, i.e., $H_0: \Lambda_k = \Lambda_{k+1}$. Permutation test is a non-parametric hypothesis test \citep{good1994} and is commonly used to assess the statistical significance when the distribution of the test statistic is not known and needs to be empirically derived. Here, we introduce essential notations for Algorithm 1. 

For all $1\leq k\leq \ell$, define $\widetilde{\mathcal{I}}_k'$ to be the $\widetilde{\mathcal{I}}$ after removing taxa with degree less than $q_k$, and let $r_k$ be the number of taxa removed. Let $\Delta_{k+1}=\Lambda_{k+1}-\Lambda_k$. If $D$ is any subset of the columns of the full OTU table, define $\widetilde{\mathcal{I}}_D$ as the adjusted mutual information matrix of $D$.\\
 
\begin{algorithm}[H]
\SetAlgoLined
{\nonl\textbf{Input}: significance level $\alpha$, number of permutations $M$, quantile vector $\boldsymbol{q}= (q_1,\cdots, q_{\ell})$.\\
\textbf{Define} :  $\boldsymbol{\hat{\theta}}=(\Delta_{k+1})\quad\text{for}\quad 1\leq k\leq \ell-1;\, \boldsymbol{\Theta}= [ \boldsymbol{\hat{\theta}}_1^*, \boldsymbol{\hat{\theta}}_2^*, \cdots,
\boldsymbol{\hat{\theta}}_M^*]$ }.  \\
Calculate  $\|\widetilde{\mathcal{I}}\|_F^2$.\\
Calculate $\boldsymbol{\hat{\theta}}$.\\
For permutation $m=1,\cdots,M$
   { \begin{enumerate}
        \item Randomly shuffle columns of taxonomy count, call this matrix $D$.
        \item For each k:
\begin{enumerate}
        \item Remove the first $r_k$ columns from $D$.
        \item Compute $\Lambda_k$ using $\widetilde{\mathcal{I}}_D$.
   \end{enumerate}
        \item Calculate the  $m$-th column of $\boldsymbol{\Theta}$, such that $\boldsymbol{\hat{\theta}}_{m}^* = (\Delta_{k+1}).$        
    \end{enumerate}}
    For each $k$, compute $p_k =\frac{\#\boldsymbol{\Theta}_k\geq\boldsymbol{\hat{\theta_k}}}{M}$, where $\boldsymbol{\Theta}_k$ is the $k$-th row of $\boldsymbol{\Theta}$. \\
 Calculate p-values: $\boldsymbol{p}= (p_k)$.\\
 Find the index of the first entry of $\boldsymbol{p}$ that is less than or equal to $\alpha$. Call this index $ind$.
\\
Remove all taxa with degree less than $q_{ind-1}$.\\
    \label{Alg:one}
\caption{ \textsc{Algorithm 1}}
\end{algorithm}

\subsubsection{\textbf{Hypothesis Testing Using Bootstrap}}
In previous subsection we described a permutation test as a useful hypothesis testing tool. Here we use bootstrap methods \citep{efron1994}, Algorithm 2, to test the same hypothesis. Again, we specifically wish to test $H_0: \Lambda_{k}=\Lambda_{k+1}$. Similar to permutation tests, a bootstrap hypothesis test is based on a test statistic.
Here, we introduce essential notations for Algorithm 2.  Let $q_1, \cdots, q_{\ell}$ be quantile values. Let $X = (x_{ij})$ for $1\leq i\leq n$ and $1\leq j\leq m$ be the taxa count matrix. For all $1\leq k\leq \ell$, define $\tilde{\mathcal{I}}_k'$ to be the columns of $\tilde{\mathcal{I}}$ after removing taxa with degree less than $q_k$. Define $X_k$ to be the subset of the columns of $X$ corresponding to the columns of $\tilde{\mathcal{I}}_k'$. Let $\Sigma_k$ be the covariance matrix of $X_k$ and $m_k$ be the number of taxa in $X_k$.
Consider the test statistic
\begin{equation}
    t_k=\frac{\Lambda_{k+1}-\Lambda_k}{\sqrt{\|\Sigma_{k+1}\|^2/m_{k+1}+\|\Sigma_{k}\|^2/m_{k}}}. \label{eq:tk}
\end{equation} 
We next describe our bootstrap process and bootstrap test statistic $t^*$. For each $k$ and $b=1,\dots,B$, sample $m_k+m_{k+1}$ columns with replacement from $(X_{k}, X_{k+1})$ and name this matrix $\boldsymbol{N}^{*}$. In addition,  We denote the first $m_{k}$ columns of $\boldsymbol{N}^{*}$, $\boldsymbol{Z}^{*}$ and the remaining $m_{k+1}$ columns $\boldsymbol{Y}^{*}$. Let $\Sigma^{*}(\boldsymbol{Z}^{*})$ ($\tilde{\mathcal{I}}'(\boldsymbol{Z}^{*})$) and $\Sigma^{*}(\boldsymbol{Y}^{*})$ ($\tilde{\mathcal{I}}'(\boldsymbol{Y}^{*})$) be the covariance (adjusted mutual information) matrices of $\boldsymbol{Z}^{*}$ and $\boldsymbol{Y}^{*}$, respectively. Let $\tilde{\mathcal{I}}'(\boldsymbol{N}^{*})$ be the adjusted mutual information matrix of $\boldsymbol{N}^*$ and define, 
\begin{equation}
\Lambda^{*}(\boldsymbol{Z}^{*}) = 1-\frac{\Vert \tilde{\mathcal{I}}'(\boldsymbol{Z}^{*}) \Vert_F^2}{\Vert \tilde{\mathcal{I}}(\boldsymbol{N}^{*}) \Vert_F^2}\ \ \ \ \text{and} \ \ \ \ \hfill\Lambda^{*}(\boldsymbol{Y}^{*}) = 1-\frac{\Vert \tilde{\mathcal{I}}'(\boldsymbol{Y}^{*}) \Vert_F^2}{\Vert \tilde{\mathcal{I}}(\boldsymbol{N}^{*}) \Vert_F^2}.\label{eq:lambdastar}
\end{equation}
Lastly, define our bootstrap test statistic to be
\begin{equation}
t_{kb}^{*} = \frac{\Lambda^{*}(\boldsymbol{Y}^*) - \Lambda^{*}(\boldsymbol{Z}^*) - (\Lambda_{k+1} -\Lambda_{k})}{\sqrt{\Vert\Sigma^*(\boldsymbol{Y}^*)\Vert^2/m_{k+1}+\Vert\Sigma^*(\boldsymbol{Z}^*)\Vert^2/m_k}}.\label{eq:tkstar}
\end{equation}

\begin{algorithm}[H]
\SetAlgoLined
{\nonl \textbf{Input}: significance level $\alpha$, the number of bootstrap samples $B$, quantile vector $\boldsymbol{q}=(q_1,\cdots, q_{\ell})$}\\
For each k:
{\begin{enumerate}
    \item Calculate $t_k$ (Eq. \ref{eq:tk}).
    \item for each $b=1,\cdots, B$:
    {\begin{enumerate}
    \item Generate $\tilde{\mathcal{I}}(\boldsymbol{N}^*)$, $\tilde{\mathcal{I}}'(\boldsymbol{Z}^*)$, $\tilde{\mathcal{I}}'(\boldsymbol{Y}^*)$, and Calculate $t_{kb}^*$ (Eq. \ref{eq:tkstar}).

    \end{enumerate}}
    \item Compute $Pval_k =\frac{\#t_{kb}^*\geq t_k}{B}$, \\
\end{enumerate}{}}
$\boldsymbol{Pval}=$($Pval_1,\cdots,Pval_{\ell}$).\\
 Find the index of the first entry of $\boldsymbol{Pval}$ that is less than or equal to $\alpha$. Call this index $ind$.\\
Remove all taxa with total abundance less than $q_{ind-1}$.\\
\label{Alg:2}
\caption{ \textsc{Algorithm 2}}
\end{algorithm}

\section{Evaluating Filtering Method}
\subsection{\textbf{Mock Microbial Community}}
To test our method, we used a publicly available mock community data set given in \citet{Brooks2015} where the ground-truth was known. These data consist of prescribed proportions of cells from seven vaginally-relevant bacterial strains: \textit{Atopobium vaginae, Gardnerella vaginalis, Lactobacillus crispatus, Lactobacillus iners, Prevotella bivia, Sneathia amnii}, and \textit{Streptococcus agalactiae} to quantify and characterize bias introduced in the sample processing pipeline such as DNA extraction, PCR amplification, and sequencing classification. 

\begin{figure}[p]
\centering
\subfloat[]{%
\resizebox*{7cm}{!}{\includegraphics{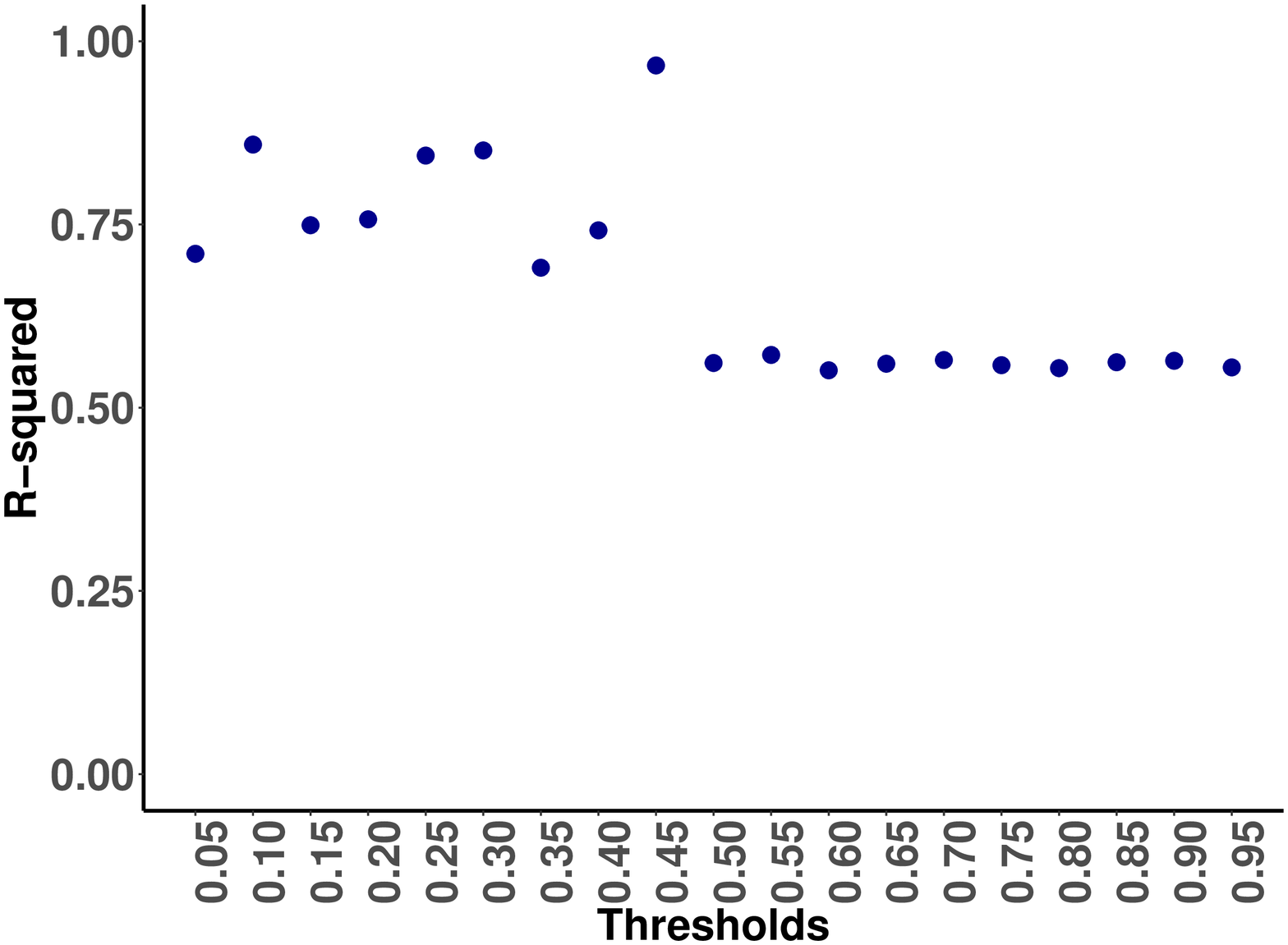}}}\hspace{5pt}
\subfloat[]{%
\resizebox*{7cm}{!}{\includegraphics{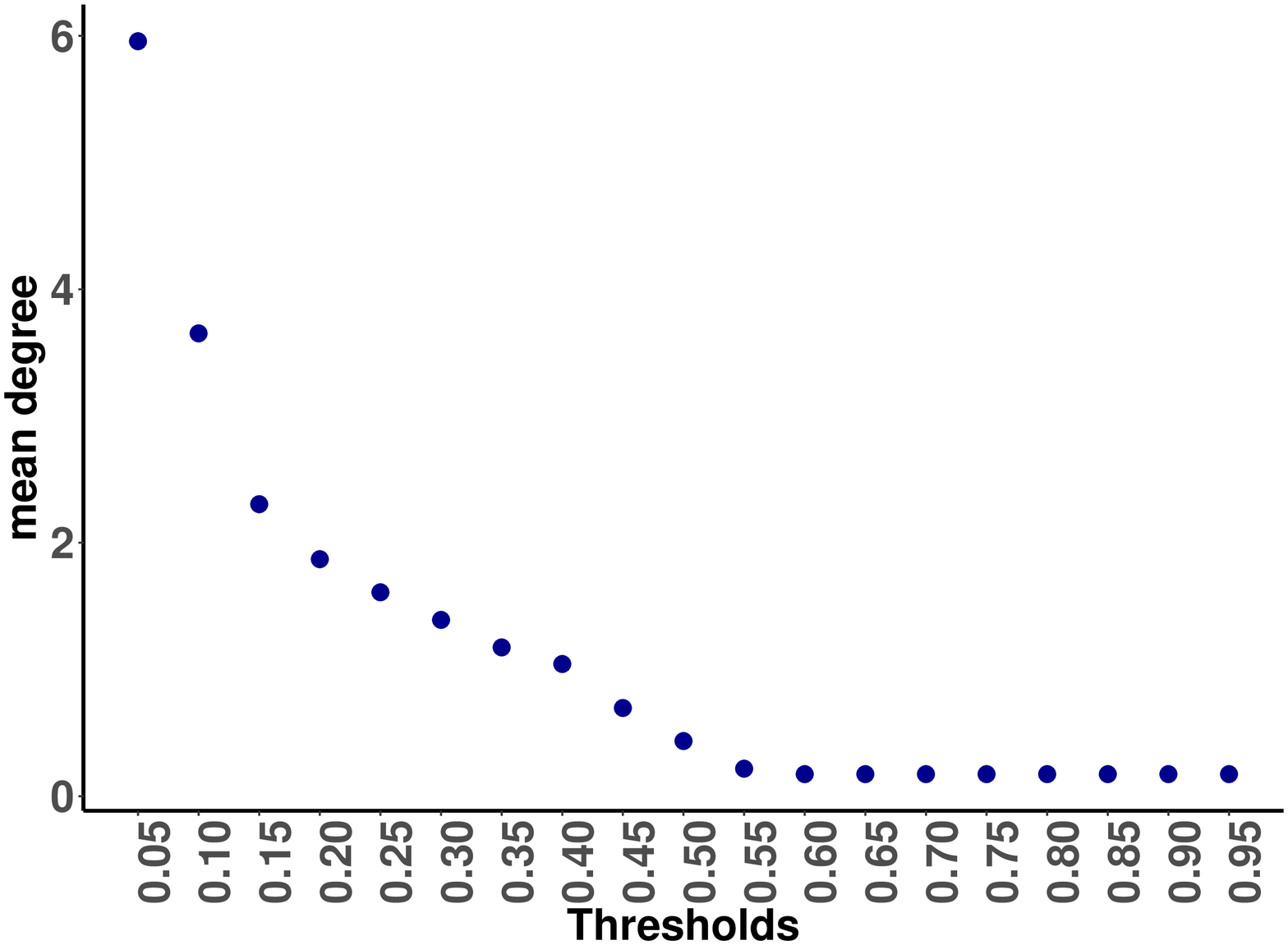}}}
\caption{Selecting a hard-threshold based on the $R^2$ (a) and the mean degree values (b) plotted versus hard-thresholding values. The hard-thresholded value at 0.45 maximizes scale-free topology and levels off after a drop at 0.5. The corresponding mean degree values are relatively low indicating a sparsity of the underlying graph. The numbers on the x-axis represent different thresholds which are plotted for illustration purposes.} \label{fig:1}
\end{figure}

We start by constructing an unweighted network of vaginal microbiome data. Table \ref{table:1} and Figure \ref{fig:1} report the results for varying the threshold parameter $\tau$ for the mock community data. It can be seen that the coefficient of determination $R^2 = 0.97$ clearly favors $\tau = 0.45$. Based on these results,  we use $\tau=0.45$ to construct the unweighted network. 

\begin{figure}[p]
\centering
\subfloat[]{%
\resizebox*{7cm}{!}{\includegraphics{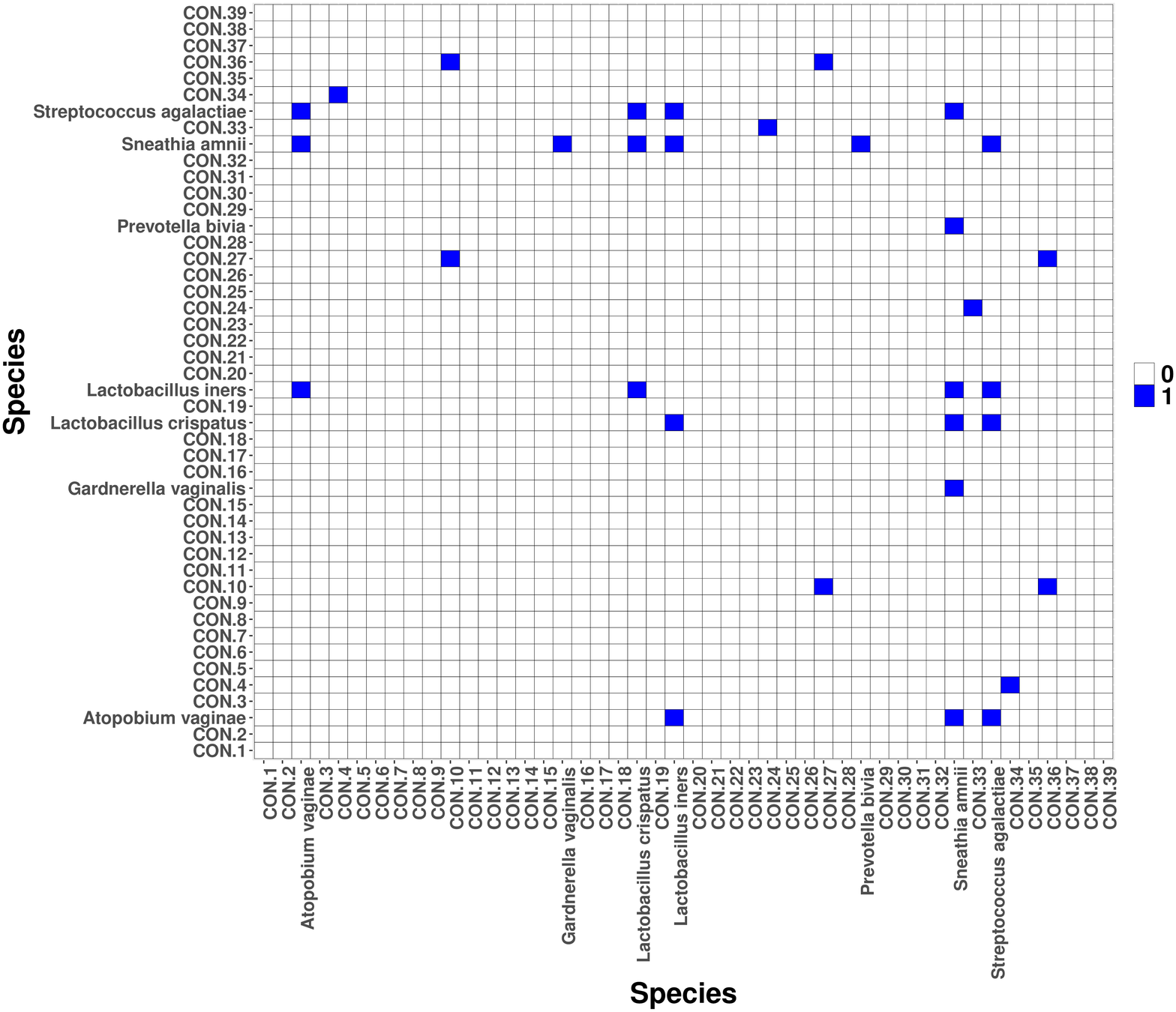}}}\hspace{5pt}
\subfloat[]{%
\resizebox*{7cm}{!}{\includegraphics{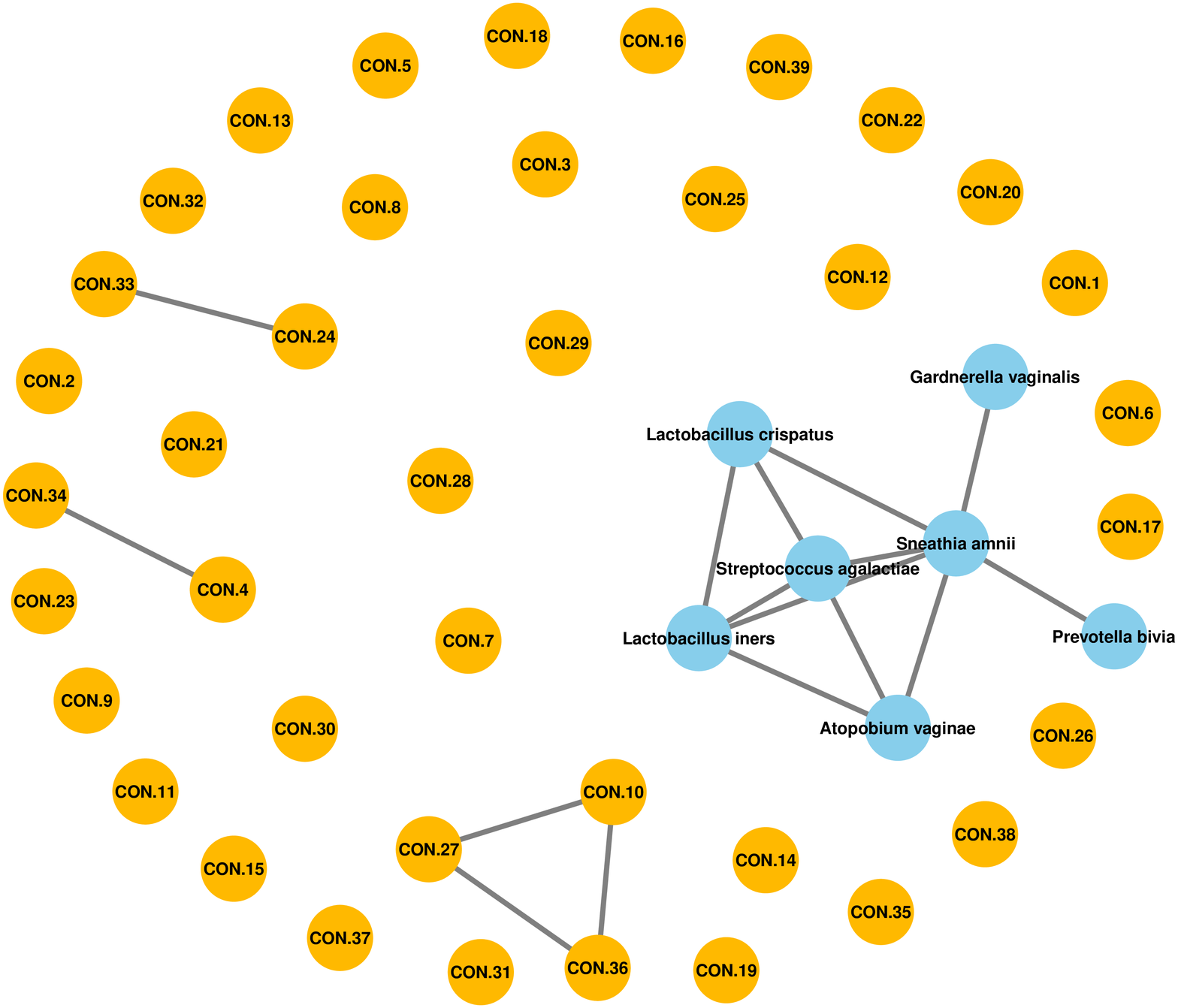}}}
\caption{Schematic diagram of an unweighted microbiome network based on adjusted mutual information. (a) adjacency matrix with $\tau=0.45$ threshold; (b) the microbiome network diagram was formed according to the relationship among 46 taxa. we indicate  contaminant taxa as CON.(arbitrary number) for convenience in illustrative purpose.} \label{fig:2}
\end{figure}

\begin{figure}[p]
\centering
\subfloat[]{%
\resizebox*{7cm}{!}{\includegraphics{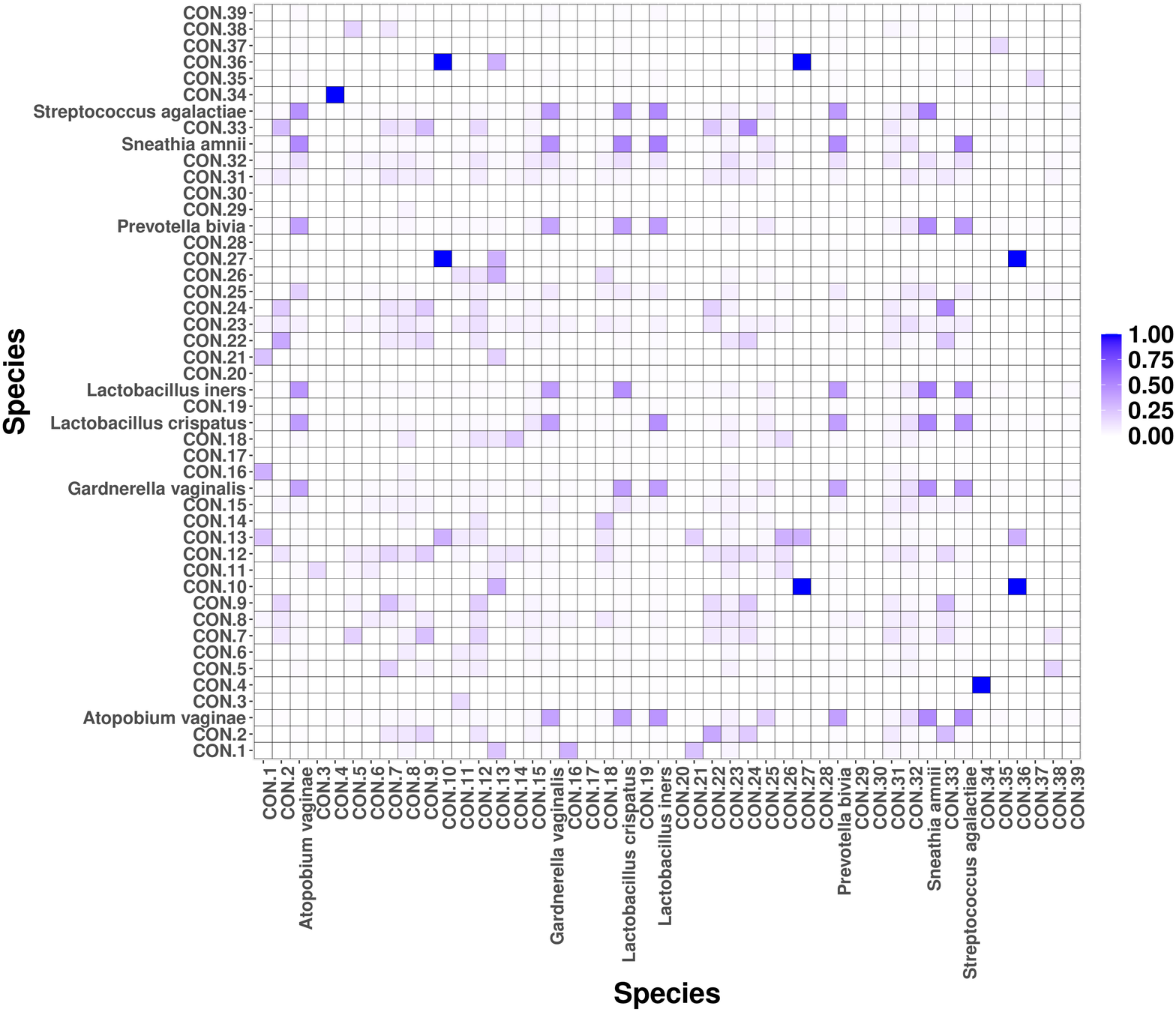}}}\hspace{5pt}
\subfloat[]{%
\resizebox*{7cm}{!}{\includegraphics{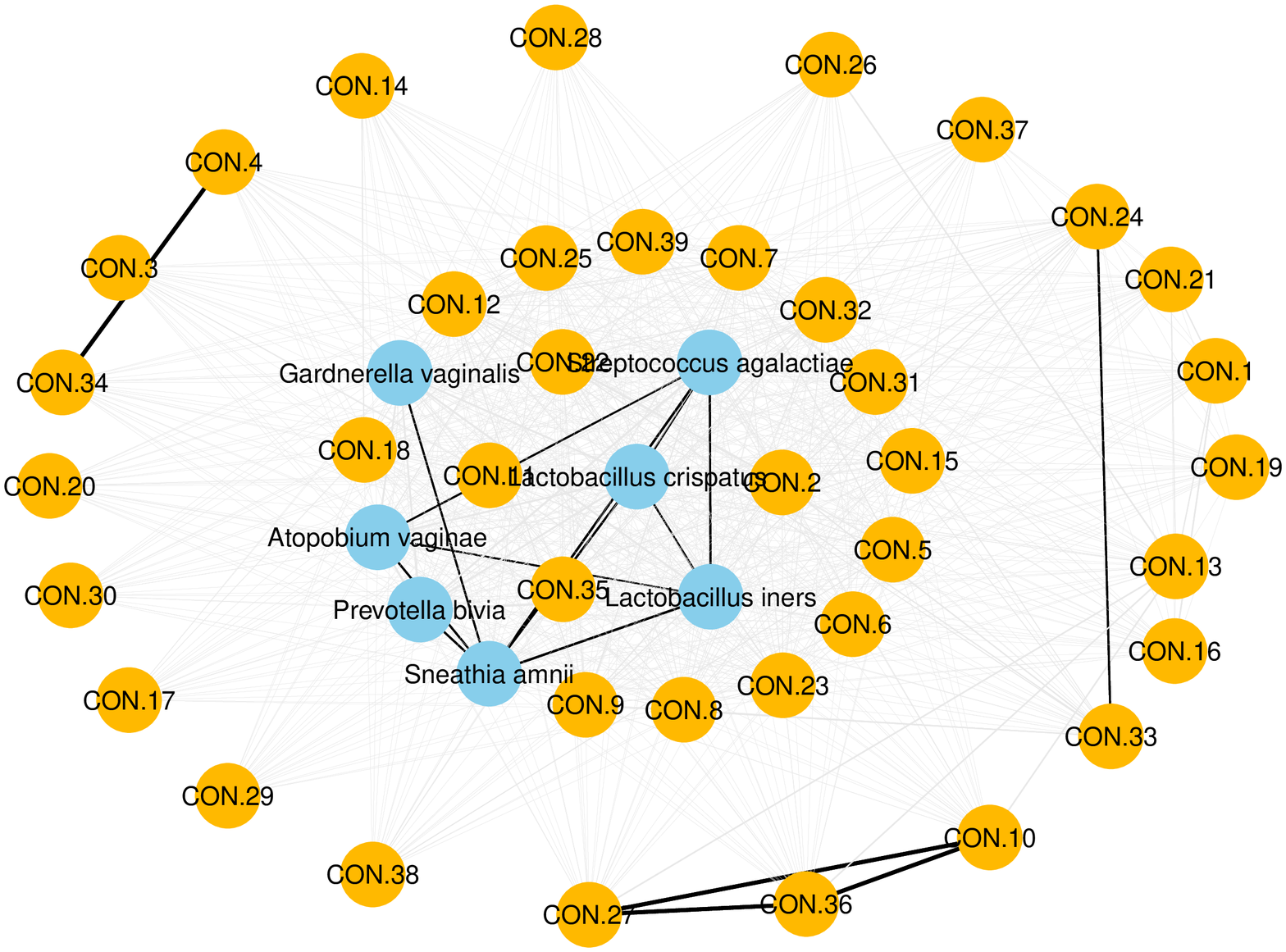}}}
\caption{Schematic diagram of a weighted microbiome network based on adjusted mutual information. (a) adjacency matrix;  (b) the microbiome network diagram was formed according to the relationship among 46 taxa. we indicate  contaminant taxa as CON.(arbitrary number) for convenience in illustrative purpose.} \label{fig:3}
\end{figure}

In Figure \ref{fig:2} we have established the adjusted mutual information  unweighted network of this dataset. It can be seen that $\mathcal{I}^*$ can reflect the true connection between the microbiome as a subnetwork and majority of noise taxa are indicated as isolated nodes. In addition, we can define the weighted network where weights are adjusted mutual information ($\widetilde{\mathcal{I}}$), this is shown in Figure \ref{fig:3}. It can be seen that in these types of networks all the nodes are connected to all other nodes. Here, edges are colored based on the strength of the connectivity between adjacent nodes from very weak (light grey), moderate (grey), strong (black). Notice that the weight between majority of true taxa is strong, however we can see three subnetworks of noise that strongly share information.

\subsubsection{Receiver Operator characteristic (ROC)}
Here we use ROC curve to evaluate the classification accuracy of each taxon in this data set using a thresholding parameter $\tau$ in reference to the binary outcome $D$, which takes $0$ (noise taxon) or $1$ (true taxon). In order to do this, we measure the degree $d_j$ of each node (taxon) in our unweighted network obtained by different hard thresholding parameter $\tau$. For each taxon, convention dictates that a true taxon is defined as $d_j\geq \tau$. The classification accuracy of each taxon is then evaluated by considering a confusion matrix. It cross-classifies the predicted outcome for taxon with $d_j\geq \tau$ versus the true outcome $D$. For the fixed cutoff $\tau$, the true positive fraction is the probability of identifying a taxon as a true signal, when it is truly a taxon.

In general, ROC analysis assesses the trade-offs between the test's fraction of true positives versus the false positives as $\tau$ varies over the range of $0$ to $1$.

\[\text{TPF}(\tau)=P(d_j\geq \tau| D=1),\]
and the false positive fraction is the probability of identifying a taxon as a true signal, while it is a noise taxon. 
\[\text{FPF}(\tau) = P(d_j\geq \tau| D=0).\]

\begin{figure}[p]
\centering
{\resizebox*{8cm}{!}{\includegraphics{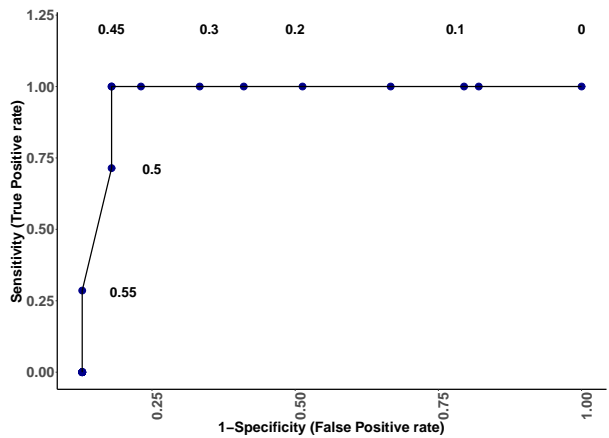}}
\caption{ROC curve describing true positive versus false positive rate of the unweighted adjusted mutual information network model predicting true taxa for \cite{Brooks2015} data.} \label{fig:4}}
\end{figure}

Since $\tau$ is not fixed in advance, one can plot TPF (sensitivity) y-axis against $\text{FPF}$(1-specificity) x-axis for all possible values of $\tau$.  If $\text{TPF}(\tau)= \text{FPF}(\tau)$, for all $\tau$, it is a useless test regards to the binary prediction. A perfect test that is completely informative about the signal status has $\text{TPF}(\tau)=1$ and $\text{FPF}(\tau)=0$ for at least one value $\tau$.  In other words, an excellent model has area under the ROC curve near $1$ which indicates a good measure of separability. A model with area near $0$ indicates a good measure of separability but a poor classification accuracy. An area under the ROC curve of $0.5$ means model has no class separation ability and is considered to be a random classifier.  
Figure \ref{fig:4} shows the ROC curve, assessing true versus false positive rate with AUC = 0.86 demonstrates the good performance of the method. 

\begin{table}[ht]
\centering
\begin{tabular}{|l||c|c|c|}
\textbf{m=46} &Predicted: True Taxa & Predicted:  Noise Taxa \\ 
  \hline
Actual True   & True Positive & False Negative \\ 
\end{tabular}
\end{table}

\begin{figure}[p]
\centering
\subfloat[]{%
\resizebox*{7cm}{!}{\includegraphics{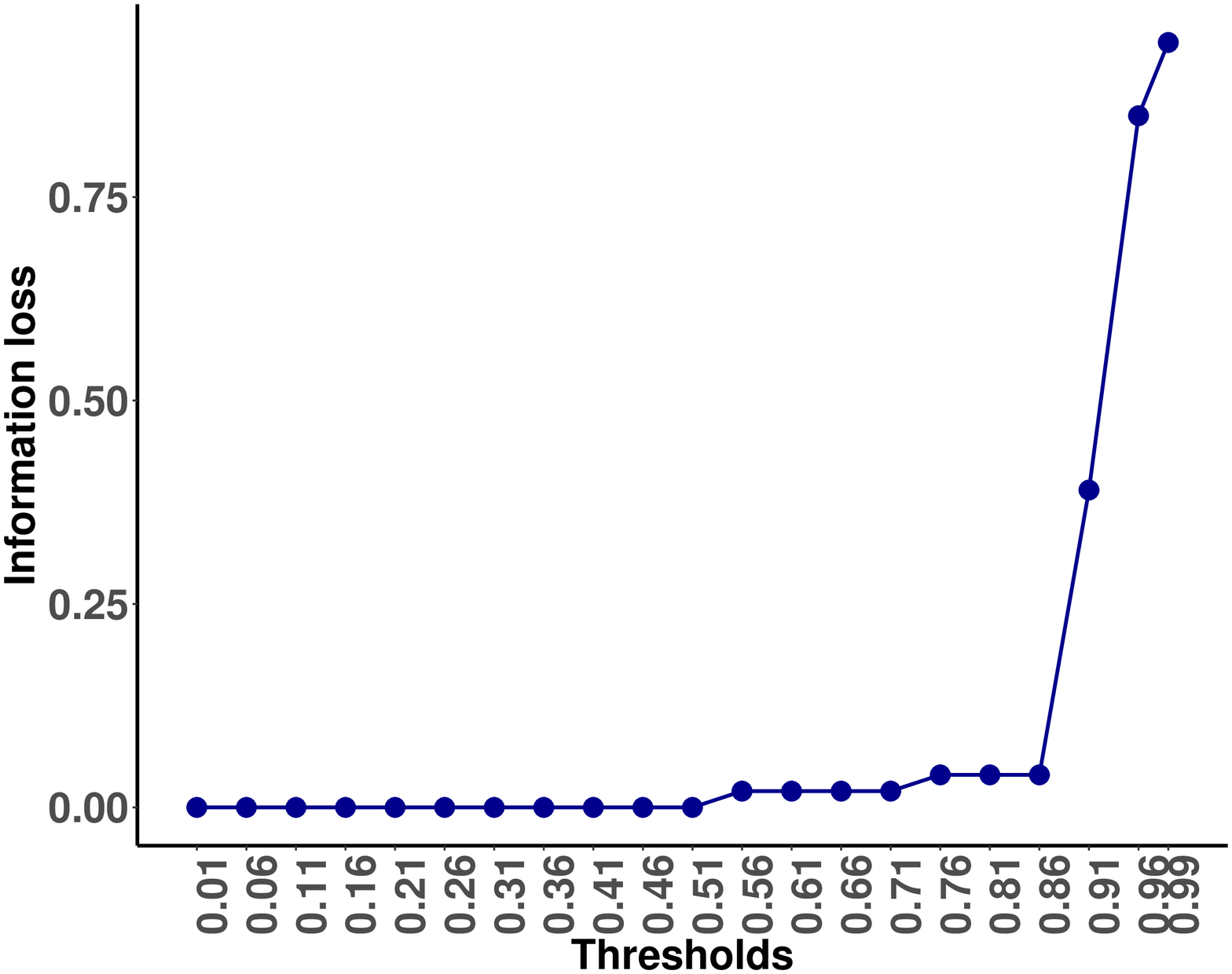}}}\hspace{5pt}
\subfloat[]{%
\resizebox*{7cm}{!}{\includegraphics{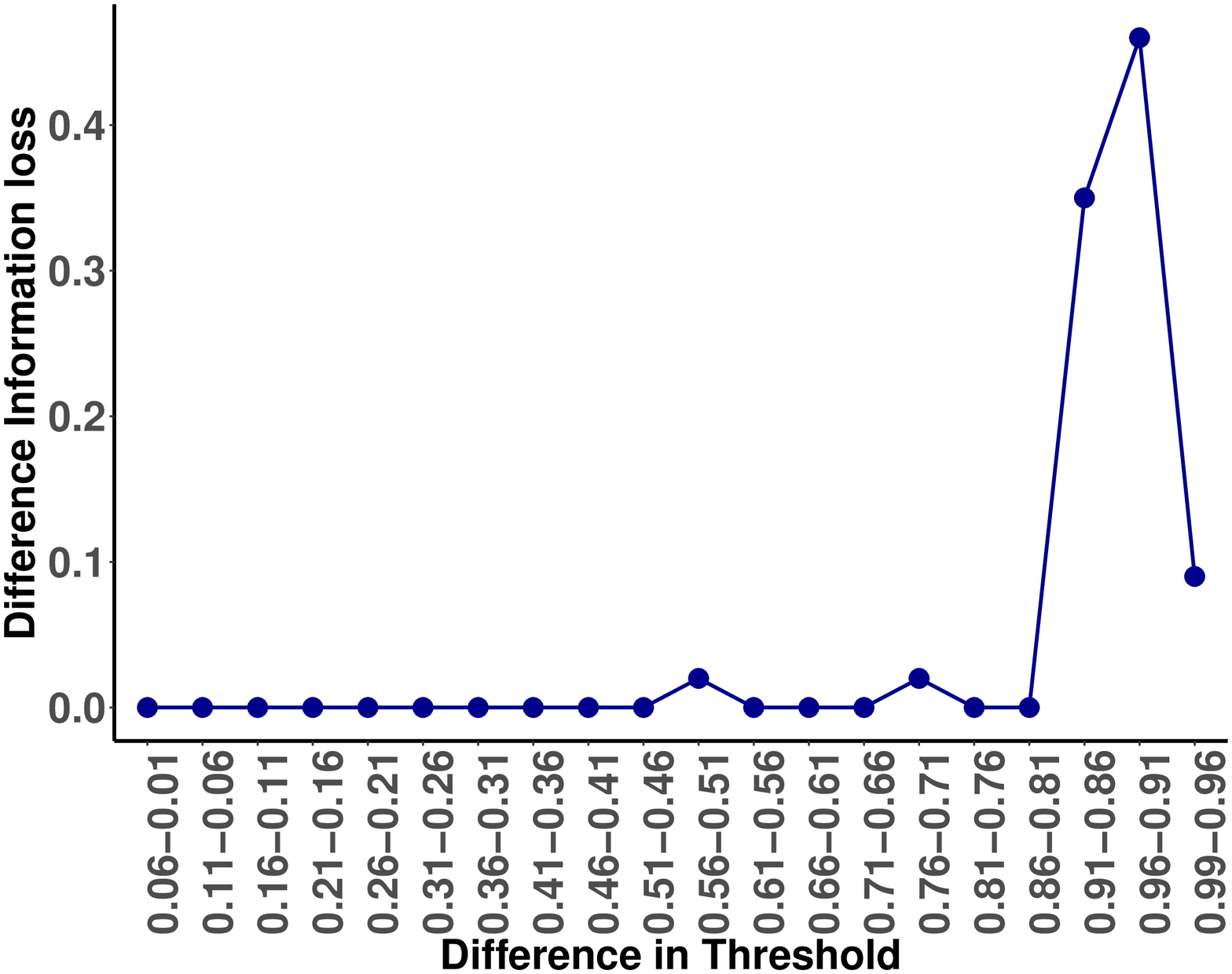}}}
\caption{Information loss for \cite{Brooks2015} data. (a) Information loss as a function of threshold. Taxa are sorted according to the increasing connectivity degree of taxa and are removed based on different percentiles. (b) Difference in information loss that evaluates the slope at each taxon.} \label{fig:5}
\end{figure}

We use  \cite{Brooks2015} data set to estimate the amount of information loss by removing different percentages of taxa and investigate if the difference in information loss by removing percentages of taxa with degree less than $q_k$ versus $q_{k+1}$ is significant. Figure \ref{fig:5} illustrates the results for this mock community data set.The left panel displays the information loss and the right panel displays the difference in information loss.  The data set was sorted according to the increasing connectivity degree of taxa using adjusted mutual information adjacency matrix. For example, a cutoff assignment of $1\%$ removes $1\%$ of the Taxa with the lowest connectivity degree. It is clear that applying percentile based filtering changes the amount of information loss, Figure \ref{fig:5}. For example,  information loss has a drastic increase from $0.86$ to $0.91$ filtering threshold, while there is  no or minimal change in information loss between removing $81\%$ and $86\%$ of taxa. This provides us with the intuition that $86\%$ of taxa can be removed from the further analysis without loosing significant amount of information and these taxa could be the result of sequencing or PCR error, especially in high-throughput sequencing data sets.

\begin{figure}[p]
\centering
{\resizebox*{8cm}{!}{\includegraphics[width=0.55\textwidth]{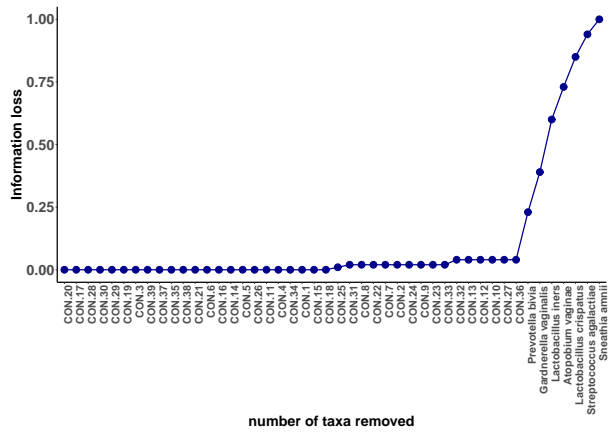}}
\caption{Information loss for \cite{Brooks2015} data. Information loss as a function of number of taxa being removed. Taxa are sorted according to the increasing connectivity degree of taxa and are removed one at a time. we indicate  contaminant taxa as $\text{CON.}\#$ for convenience in illustrative purpose.}
\label{fig:6}}
\end{figure}

Figure \ref{fig:6} shows information loss versus the number of taxa that are removed based on the lowest connectivity degree one at a time. We can see that after removing true signals (indicated by taxonomic name) the information loss values increase dramatically. From the Figures \ref{fig:5} and \ref{fig:6}, it is clear that information loss increases after removing a certain number or percentage of taxa. However, we need to investigate whether this rise of information loss is due to random errors or a real effect. In other word, we want to determine whether the  information loss after removing less than $q_k$ of taxa is significantly different from information loss after removing $q_{k+1}$ of taxa, where $q_k$ is the $k-th$ quantile value of connectivity degree. To do this, we use permutation and bootstrapping approaches to test the null hypothesis which indicates that removing taxa with degree less than $q_{k+1}$ versus $q_{k}$ does not make any difference in information loss and hence we can remove them from further analysis. We follow the Algorithm 1 and 2 by setting  $M=500$, $B=500$, $\alpha=0.05$, and $\mathbf{q} = (0.01,\cdots,0.96)$. The results of permutation and bootstrapping tests are shown in Table \ref{table:2}, second column shows that there is a significant loss of information after removing $\geq 91\%$  of taxa (p-value$<0.05$) at $5\%$ significance level. As an alternative, we follow Algorithm 2 to apply bootstrap method for hypothesis testing to approximate p-value. Table \ref{table:2}, third column shows that the bootstrap test gives similar results to permutation tests which also indicates that there is strong evidence that removing $91\%$ of taxa will result in loosing significant amount of information. 

\section{Comparison Study on Mock Community Data \cite{Brooks2015}}\label{section:5}
Here, we use  data set in \cite{Brooks2015} to assess the performance of our method and compare results to alternative methods. More specifically, we consider four traditional methods which have been commonly employed for filtering of microbiome data: (1) Traditional 1: we retain taxa with more than $0.1\%$, $5\%$, and $1\%$ relative abundance in at least one sample \citep{Dobbler2019, Logares2014, Partula2019, ridenhour2019}. (2) Traditional 2: we retain taxa with at least 5 reads in at least 3 samples \citep{Ingham2019}. (3) Traditional 3: we retain taxa presented in more than 5 samples \citep{brigham2001benthic}. (4) Traditional 4:  we remove samples with fewer than 100 reads and taxa with fewer than 10 reads, as well as taxa which present in fewer than $1\%$ of samples \citep{Duvallet2017}. \\
 
\begin{figure}[p]
\centering
\subfloat[]{%
\resizebox*{7cm}{!}{\includegraphics{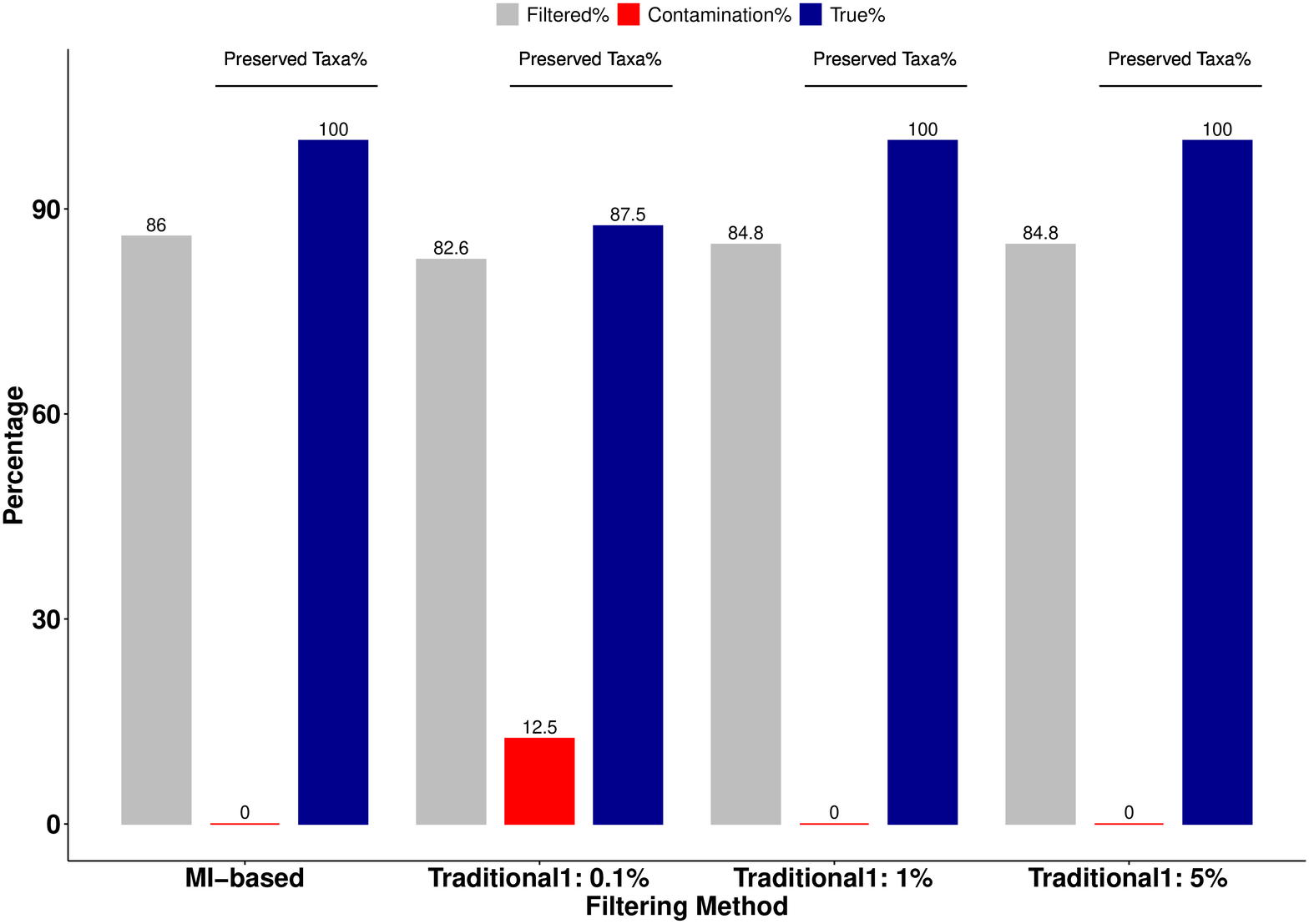}}}\hspace{5pt}
\subfloat[]{%
\resizebox*{7cm}{!}{\includegraphics{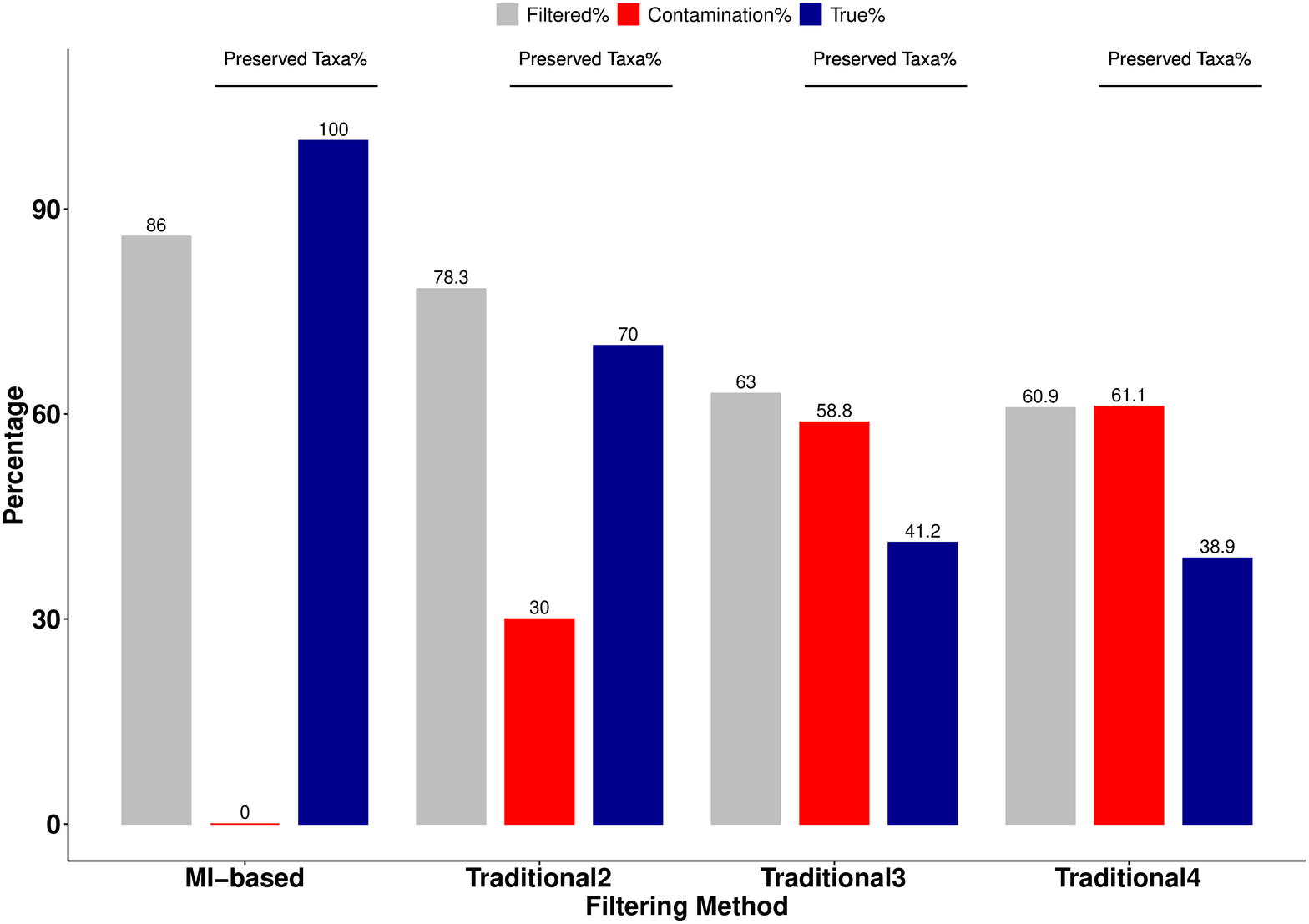}}}
\caption{Comparison of Traditional filtering methods and MI-based filtering approach using data set \cite{Brooks2015}. (a) Bar graph showing Traditional 1 with $0.1\%$, $1\%$, and $5\%$ removal thresholds and MI-based filtering methods. (b) Traditional 2, 3, 4 and MI-based filtering methods. The colored bars represent: percentage of filtered taxa in each method (grey), percentage of contamination preserved after filtering (red), percentage of true taxa preserved after filtering (blue).} \label{fig:7}
\end{figure}

 Results presented in Figure \ref{fig:7} and Table \ref{table:3} indicated that MI-based filtering method performs better than Traditional 2, 3, and 4 as well as Traditional 1 for the choice of $0.1\%$. In particular, we can see than MI-based method removed $86\%$ of taxa with minimum loss of information and preserved $14\%$ of taxa which were all true signals ($100\%$). Traditional 1 filtering method with retain threshold of $1\%$ and $5\%$ performed as good as MI-based method, however, when retain threshold was $0.1\%$ this method retained $12.5\%$ of contamination. In Traditional 2 filtering method, following filtering $78\%$ of taxa, this method retained $30\%$ noise signals, Similarly in Traditional 3 and 4, they retained around $60\%$ and $61\%$ non-bacterial signals, respectively. Therefore, proposed method was successfully able to identify and remove contaminants from this data which result in dimensionality reduction of the data that can reduce computation time and improve interpretation of the downstream analysis. The proposed method has two advantages in addition to its superior performance in comparison with above mentioned traditional methods. First, it does not required a choice of threshold which is critical and not easy to obtain. Second, it is able to detect true taxa with low abundance. Most of the traditional methods have subjective predetermined thresholding value that might have adverse effects on the analysis due to loss of important information within filtered taxa. Our proposed method choose filtering threshold based on hypothesis testing and information loss. As mentioned earlier, these traditional methods remove taxa with low abundance and hence any important taxa with low abundance is removed leading to significant loss of information. However, MI-based filtering method, removes taxa based on their interactions with other taxa and therefore it can preserve low abundance taxa in case of their strong association with other taxa. This allows us to study significant taxa that occur in low abundance. 
 
\section{Conclusion}
Removing contaminants prior to any downstream analysis is an essential step in metagenomic sequencing data research. Host associated contaminants significantly complicate analysis, particularly in low microbial biomass body sites. Contamination can cause analysis of sequencing reads to result in false positive or false negative and hence decreasing the reliability of the analysis. Here, we developed a simple  method that uses combination of graph theory and information theoretic functionals to identify and remove contaminants in metagenomic data sets. Our results suggest that mutual information based filtering method can improve the accuracy of detecting  contaminants, especially in comparison with the commonly used traditional filtering methods. 

To fully explore the strengths and weaknesses of our proposed filtering method, evaluation on different labeled mock community data sets are necessary. Unfortunately, labeled data sets are expensive and difficult to obtain and hindered our ability to test our method further. We believe that it is possible to improve the threshold selection in the unweighted graph given improved (ground-truthed) data with which to work. Looking solely for isolated nodes is not sufficient as it is unable to filter out random interaction between contaminants nodes. When the number of nodes increases we can expect more random interaction between contaminants, making the isolated node approach even less powerful. We believe more advanced methods from graph theory could remedy this short coming. Future work could include but not limited to examination the efficiency of techniques such as dense community detection, dense subgraph selection, and vertex selection based on vertex centrality. These sophisticated node selection methods could provide a more powerful filtering method in the unweighted graph.

%

%
%
%
%

%

\begin{table}[p]
\caption{Microbiome network characteristics for different hard threshold $\tau$. The second column contains coefficient of variation $R^2$ that varies between $0$ and $1$, where $0$ indicates that the power-law model explains none of the variability of the empirical degrees, while $1$ indicates the model perfectly fit the data.}
\centering
\begin{tabular}{||c||c|c|c|c|c|}
  \hline
$\tau$ & $R^2$ & $\bar{k}$ & $-\gamma$ \\ 
  \hline
 0.05 & 0.71 & 5.96 & -0.85 \\ 
 0.10 & 0.86 & 3.65 & -1.00 \\ 
 0.15 & 0.75 & 2.30 & -0.56 \\ 
 0.20 & 0.75 & 1.87 & -0.64 \\ 
 0.25 & 0.84 & 1.61 & -0.60 \\ 
 0.30 & 0.84 & 1.39 & -0.61 \\ 
 0.35 & 0.70 & 1.17 & -2.68 \\ 
0.40 & 0.74 & 1.04 & -0.69 \\ 
\boxit{1.7in} \textbf{0.45} & \textbf{0.97} & \textbf{0.70} & \textbf{-0.97} \\ 
 0.50 & 0.56 & 0.43 & -4.53 \\ 
 0.55 & 0.56 & 0.22 & -3.90 \\ 
 0.60 & 0.57 & 0.17 & -3.96 \\ 
 0.65 & 0.56 & 0.17 & -3.96 \\ 
 0.70 & 0.56 & 0.17 & -3.96 \\ 
 0.75 & 0.56 & 0.17 & -3.96 \\ 
 0.80 & 0.56 & 0.17 & -3.96 \\ 
 0.85 & 0.56 & 0.17 & -3.96 \\ 
 0.90 & 0.56 & 0.17 & -3.96 \\ 
 0.95 & 0.57 & 0.17 & -3.96 \\ 
   \hline
\end{tabular}
\label{table:1}
\end{table}

\clearpage 

\begin{table}[p]
\caption{P-values by permutation and bootstrapping test}
\centering
\begin{tabular}{|c|c|c|}
  \hline
 Percentage of taxa removal& p-value by permutation & p-value by bootstrapping \\ 
  \hline
  0.01-0.06 & $>0.1$& 0.508 \\ 
  0.06-0.11 & $>0.1$& 0.514\\ 
  0.11-0.16 & $>0.1$ &0.503\\ 
  0.16-0.21 & $>0.1$ &0.491\\ 
  0.21-0.26 & $>0.1$ &0.511\\ 
  0.26-0.31 & $>0.1$ &0.575\\ 
  0.31-0.36 & $>0.1$ &0.644\\ 
  0.36-0.41 & $>0.1$&0.525 \\ 
  0.41-0.46 & $>0.1$ &0.601\\ 
  0.46-0.51 & $>0.1$ &0.663\\ 
  0.51-0.56 & $>0.1$ &0.507\\ 
  0.56-0.61 & $>0.1$ &0.585\\ 
  0.61-0.66 & $>0.1$ &0.666\\ 
  0.66-0.71 & $>0.1$ &0.534\\ 
  0.71-0.76 & $>0.1$ &0.722\\ 
  0.76-0.81 & $>0.1$ &0.753\\ 
  0.81-0.86 & $>0.1$ &0.813\\ 
\textbf{0.86-0.91} & $\mathbf{<0.05}$ &$\mathbf{<0.05}$\\ 
\textbf{0.91-0.96} & $\mathbf{<0.05}$ &$\mathbf{<0.05}$\\
&&\\
   \hline
\end{tabular}
\label{table:2}
\end{table}

\clearpage

\begin{table}[p]
\caption{Comparison of 6 commonly used traditional filtering method with MI-based filtering method for Mock community data set in \cite{Brooks2015}.}
\centering
\begin{tabular}{|rlccc|}
  \hline
  \ &\ &\ &\ &\ \\
 & Filtering Method & Filtered\% &\shortstack{Preserved\\Contamination\%}&\shortstack{Preserved\\True\%} \\
    \ &\ &\ &\ &\ \\
  \hline
   & Traditional1-0.1\% & 0.83 & 0.12 & 0.88 \\ 
   & Traditional1-5\% & 0.85 & 0.00 & 1.00 \\ 
   & Traditional1-1\% & 0.85 & 0.00 & 1.00 \\ 
   & Traditional2 & 0.78 & 0.30 & 0.70 \\ 
   & Traditional3 & 0.63 & 0.59 & 0.41 \\ 
   & Traditional4 & 0.61 & 0.61 & 0.39 \\
   & MI-based & 0.86 & 0.00 & 1.00 \\
   \hline
\end{tabular}
\label{table:3}
\end{table}

\clearpage

\end{document}